\documentclass[11pt]{article}

\usepackage[final]{acl}

\usepackage{times}
\usepackage{latexsym}

\usepackage[T1]{fontenc}

\usepackage[utf8]{inputenc}

\usepackage{microtype}

\usepackage{inconsolata}

\usepackage{graphicx}

\usepackage{amsmath,amssymb,amsfonts}
\usepackage{algorithmic}
\usepackage{booktabs}
\usepackage{subcaption}
\usepackage{float}
\usepackage[table]{xcolor}
\usepackage[table, dvipsnames]{xcolor}
\usepackage{color}   
\usepackage{multirow}
\usepackage[most]{tcolorbox}
\usepackage{enumitem}

\usepackage{listings}
\usepackage{xcolor}

\lstdefinestyle{pythonstyle}{
    language=Python,
    basicstyle=\footnotesize\ttfamily,
    keywordstyle=\color{blue},
    commentstyle=\color{gray},
    stringstyle=\color{red!70!black},
    showstringspaces=false,
    breaklines=true,
    numbers=left,
    numberstyle=\tiny\color{gray},
    numbersep=5pt,
    frame=none,
}
\title{No Test Cases, No Problem: Distillation-Driven Code Generation for Scientific Workflows}

\author{
Siddeshwar Raghavan\\
Purdue University \\
Electrical and Computer Engineering \\
West Lafayette, USA \\
\texttt{raghav12@purdue.edu} \\
\And
Tanwi Mallick \\
Argonne National Laboratory \\
Mathematics and Computer Sciences \\
Lemont, USA \\
\texttt{tmallick@anl.gov}
}

\begin{document}
\maketitle
\begin{abstract}
Existing multi-agent Large Language Model (LLM) frameworks for code generation typically use execution feedback and improve iteratively using Input/Output (I/O) test cases. However, this does not work for scientific workflows, where I/O test cases do not exist, and generating them requires solving the very problem at hand. To address this, we introduce MOSAIC, a training-free multi-agent framework for scientific code generation without I/O supervision. Instead of execution feedback, MOSAIC employs a student-teacher knowledge distillation framework that grounds generation through domain-specific examples and structured problem decomposition. To further mitigate hallucinations across chained subproblems, we introduce a Consolidated Context Window (CCW) for maintaining consistent reasoning across agents. Experiments on the SciCode benchmark show that MOSAIC improves accuracy, executability, and numerical precision over existing approaches while relying on lightweight models.
\end{abstract}

\section{Introduction}
\label{sec:intro}
Scientific coding plays a very crucial role in the process of modern empirical discovery. Across physics, biology, and materials science, computational problems demand numerical precision, reproducibility, and deep domain reasoning. These problems decompose naturally into chains of interdependent subproblems, and errors accumulate fast. As context grows, LLMs struggle to retain critical information and become prone to hallucinations and inconsistencies~\cite{LLM_hallucinate}. Collectively, these requirements exceed the capabilities of standalone LLMs and simple agent setups.

Existing multi-agent code generation frameworks try to handle coding complexity through planning, execution, and iterative debugging. However, they have one important hidden dependency, that is, \emph{Input/Output (I/O) test cases}. From one-shot synthesis~\cite{humaneval} and chain-of-thought prompting~\cite{cot} to more advanced multi-agent pipelines~\cite{agentcoder,mapcoder,codesim,mallick2024chatvis}, the standard approach improves the generated code by refining it with I/O samples. This type of structure works well in competitive programming settings, where test cases are already available, but it is not suitable for scientific discovery.

\begin{figure}[t]
\centering
\begin{tcolorbox}[title=Competitive coding (I/O test cases available), 
                  fonttitle=\small\bfseries,
                  colback=blue!3!white, colframe=blue!40]
\small
\textbf{Problem:} Write a function to find the sum of 
all even numbers in a list.\\[1pt]
\textbf{Test cases:}
\begin{lstlisting}[language=Python, basicstyle=\footnotesize\ttfamily, numbers=none]
assert sum_even([1, 2, 3, 4]) == 6
assert sum_even([1, 3, 5]) == 0
\end{lstlisting}
\end{tcolorbox}

\vspace{1pt}

\begin{tcolorbox}[title=Scientific coding (no I/O test cases), 
                  fonttitle=\small\bfseries,
                  colback=red!3!white, colframe=red!40]
\small
\textbf{Problem:} Write a Haldane model Hamiltonian 
on a hexagonal lattice.\\[1pt]
\textbf{Function header:}
\begin{lstlisting}[language=Python, basicstyle=\footnotesize\ttfamily, numbers=none]
def hamiltonian(kx, ky, a, t1, t2, phi, m):
\end{lstlisting}
\textbf{Background info:} Defines what a Haldane model is.
\end{tcolorbox}
\caption{\textbf{Problem Structure Differences}: General and Competitive coding benchmarks provide explicit I/O pairs for validation during code generation. Scientific coding benchmarks typically provide only a function signature and background information.}
\label{fig:io-comparison}
\end{figure}

\paragraph{The verification cycle.}
In scientific workflows, validation test cases are not available in advance as shown in Figure~\ref{fig:io-comparison}. Unlike competitive programming benchmarks, where expected outputs are fixed, scientific problems are controlled by domain-specific tolerances and numerical limits that vary from problem to problem. More importantly, constructing a valid I/O sample for these physical simulations or biological models is not straightforward, since the underlying algorithm is what we are aiming to generate. This leads to a fundamental deadlock: the code cannot be verified without knowing the answer, but if the answer is already known, then the problem is effectively solved. Prior frameworks such as MapCoder~\cite{mapcoder},
CodeSIM~\cite{codesim}, and AgentCoder~\cite{agentcoder} designed for competitive programming are fundamentally unsuitable for scientific settings, as their entire verification loop depends on I/O test cases.
\paragraph{Grounding without ground truth:} To address this issue, we propose \textbf{MOSAIC}, a fully autonomous and training-free multi-agent framework that helps grounding of scientific code at the architectural level, instead of entirely depending on execution-based feedback. Rather than verifying code against I/O test cases, MOSAIC decouples semantic grounding by generating domain-specific rationales through a student–teacher knowledge distillation framework. This process is kept separate from syntactic grounding, which is managed by a dedicated Debugger Agent that focuses on resolving syntax and import errors. Our proposed Consolidated Context Window (CCW) maintains consistent reasoning across chained subproblems without expanding the context window. We evaluate extensively on SciCode~\cite{scicode} benchmark, MOSAIC consistently outperforms all baselines across LLM backbones and scientific domains.
\noindent Our main contributions are: 
\begin{itemize} 
\item We identify the \emph{verification cycle} as the fundamental obstacle to scientific code generation and propose architectural grounding as a principled solution. 
\item We introduce MOSAIC, a training-free, LLM-agnostic multi-agent framework that operates entirely without I/O test cases, decoupling semantic and syntactic grounding. 
\item We introduce the Consolidated Context Window (CCW), which preserves reasoning coherence across interdependent subproblems without growing the context window. 
\item Extensive experiments and ablations on SciCode show the promise of our proposed approach achieving up to 24\% accuracy gains across the five scientific domains. 
\end{itemize}
\section{Related Work}
\label{sec:rel_work}

\textbf{LLMs for Science \& Code Generation:}
LLMs are increasingly adopted for scientific discovery~\cite{Wang2023ScientificDI, ai4science2023impactlargelanguagemodels, zhang2024scientificllms_survey} and code generation~\cite{humaneval, codegen, santacoder, starcoder}. While techniques like Chain-of-Thought~\cite{cot}, Self-Consistency~\cite{wang2023selfconsistencyimproveschainthought}, and Graph of Thoughts~\cite{graphoffthoughts} improve multi-step reasoning, and methods like Reflexion~\cite{reflexion} mimic the experimental trial-and-error cycle, these methods predominantly rely on execution feedback. General coding benchmarks (e.g., HumanEval~\cite{humaneval}, APPS~\cite{APPS}, MBPP~\cite{MBPP}) and repository-level tasks (SWE-bench~\cite{swebench}) provide validation test cases to guide algorithm improvement. This dependency limits their applicability in scientific domains where I/O targets cannot be trivially generated. In contrast, the SciCode benchmark~\cite{scicode} evaluates executable scientific programs without relying on I/O tests, exposing a gap in current methods.

\textbf{Agentic LLM Frameworks.}
Recent research explores collaborative multi-agent pipelines, where specialized agents interact to tackle complex programming tasks. Frameworks like MapCoder~\cite{mapcoder} and CodeSIM~\cite{codesim} employ planning, coding, and debugging agents to iteratively verify inputs and outputs. Some systems also integrate external APIs and mathematical toolkits~\cite{mathchat, toolformer, autogen}. However, while effective for standard programming tasks, these systems fail to generalize to scientific problem-solving where reasoning must be coupled with domain knowledge and where validation I/O is unavailable to correct semantic logic. Motivated by this gap, we propose a modular, training-free multi-agent framework driven by knowledge distillation, separating semantic grounding from syntactic grounding to operate without an I/O oracle.
\section{Method}
\label{sec:method}
In this section, we present our framework, MOSAIC, detailing the various agents it orchestrates as well as the underlying design choices that allow it to operate without execution feedback.

\subsection{MOSAIC Framework}
We design MOSAIC as a modular, multi-agent framework for solving complex scientific coding problems. It operates by breaking problems into smaller steps and coordinating specialized agents to produce correct and executable code. The framework is LLM-agnostic and adapts to various fields by integrating domain-specific knowledge and memory into each agent, without fine-tuning. MOSAIC consists of four main agents \textbf{Self-Reflection}, \textbf{Rationale}, \textbf{Coding}, and \textbf{Debugging}. A \textbf{Bucketing Module} first assigns each problem to the appropriate domain. While all domains share the same agent architecture, each maintains its own dedicated memory to prevent cross-domain interference.

\begin{figure*}[!h]
    \centering
    \includegraphics[width=1\linewidth]{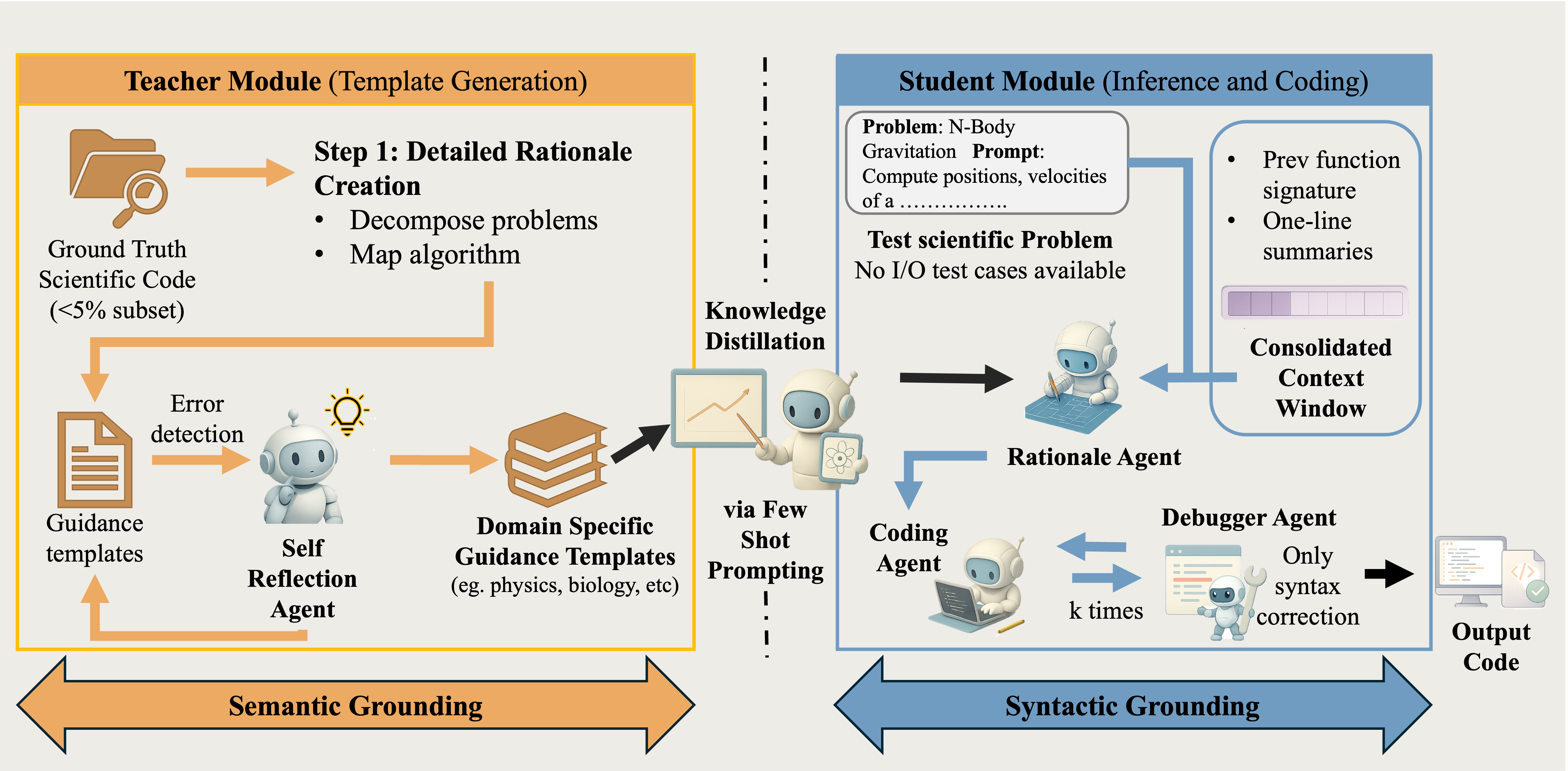}
    \caption{\textbf{MOSAIC}: our distillation-driven multi-agent framework for scientific code generation without I/O supervision. The Teacher Module derives domain-specific guidance from a small validation set and passes it to the Student Module via few-shot prompting. The Consolidated Context Window (CCW) maintains focus on the current subproblem by retaining only prior function signatures and summaries.}
    \label{fig:mosaic}
\end{figure*}

Inspired by the concept of Knowledge Distillation (KD)~\cite{hinton2015distillingknowledgeneuralnetwork}, MOSAIC is structured as a teacher-student system, as illustrated in Figure~\ref{fig:mosaic}. Within the Teacher Module, we use ground-truth scientific code from a small non-overlapping subset ($\le$ 5\%) of the validation data to create detailed domain specific rationales that break each problem into a sequence of solution steps. These rationale steps, along with the corresponding ground-truth code, enable the teacher to create domain-specific guidance templates. These templates then help the Student Module learn to solve problems within a given scientific domain. Because I/O test cases are absent, this distillation process acts as the primary mechanism for \textit{semantic grounding}. In the following sections, we explore the core agents that form the foundation of our architecture.

\subsubsection{\textbf{Self-Reflection Agent}}
The Self-Reflection Agent is the key component of our Teacher Module. To construct a strong domain guidance template, this agent receives the ground-truth rationale and learns to evaluate its own reasoning steps for each domain. It identifies potential mistakes or omissions and refines its logic iteratively before arriving at the final pseudocode. By verbalizing its thought process and critically analyzing its reasoning path, the Self-Reflection Agent establishes the semantic baseline for the framework, enhancing output reliability and correcting logical flaws without executing code.

\subsubsection{\textbf{Rationale Agent}}
The Rationale Agent is the first of three components within the Student Module. It uses the template pseudocode generated by the Self-Reflection Agent as few-shot examples to process scientific problems from the test set. It then produces a clear, step-by-step reasoning plan similar to the structured guidance offered by the Teacher Module. Scientific problems typically involve a sequence of dependent subproblems that must be addressed in a specific order. To address the risk of hallucination as the context window grows~\cite{LLM_hallucinate, banerjee2024llmshallucinateneedlive}, we implement a \textbf{Consolidated Context Window (CCW)} within the Rationale Agent. As illustrated in Figure~\ref{fig:mosaic}, to mitigate hallucination, the CCW contains only previously implemented function signatures and brief one-sentence summaries, rather than the full code and reasoning history. Thus, the CCW helps the agent remain focused on the current task while maintaining enough historical awareness to determine the logical next steps in the reasoning process.

\subsubsection{\textbf{Coding Agent}}
The Coding Agent uses the detailed plan provided by the Rationale Agent to generate the corresponding code block in Python, maintaining awareness of both the subproblem and the broader problem context with the help of the CCW. While MOSAIC is designed for I/O-free environments, the Coding Agent remains flexible, if a prompt does include I/O test cases (as seen in general-purpose coding datasets~\cite{MBPP, humaneval, APPS}), it can successfully incorporate them to improve quality of generated code.

\subsubsection{\textbf{Debugger Agent}}
The final core agent in our MOSAIC framework is the Debugger Agent, which executes the generated code and performs up to $k$ rounds of error correction in collaboration with the Coding Agent. Because scientific coding benchmarks such as SciCode~\cite{scicode} lack I/O test cases to verify algorithmic logic, the Debugger Agent is strictly constrained to \textit{syntactic grounding}. This iterative process strictly resolves syntax and import errors. By ensuring code executability, the Debugger Agent guarantees that the semantically grounded logic generated by the previous agents can be successfully evaluated.
\section{Experiments}
\label{sec:exps}

\subsection{Datasets and Comparison Approaches}
\begin{figure*}[h]
    \centering
    \includegraphics[width=1\textwidth]{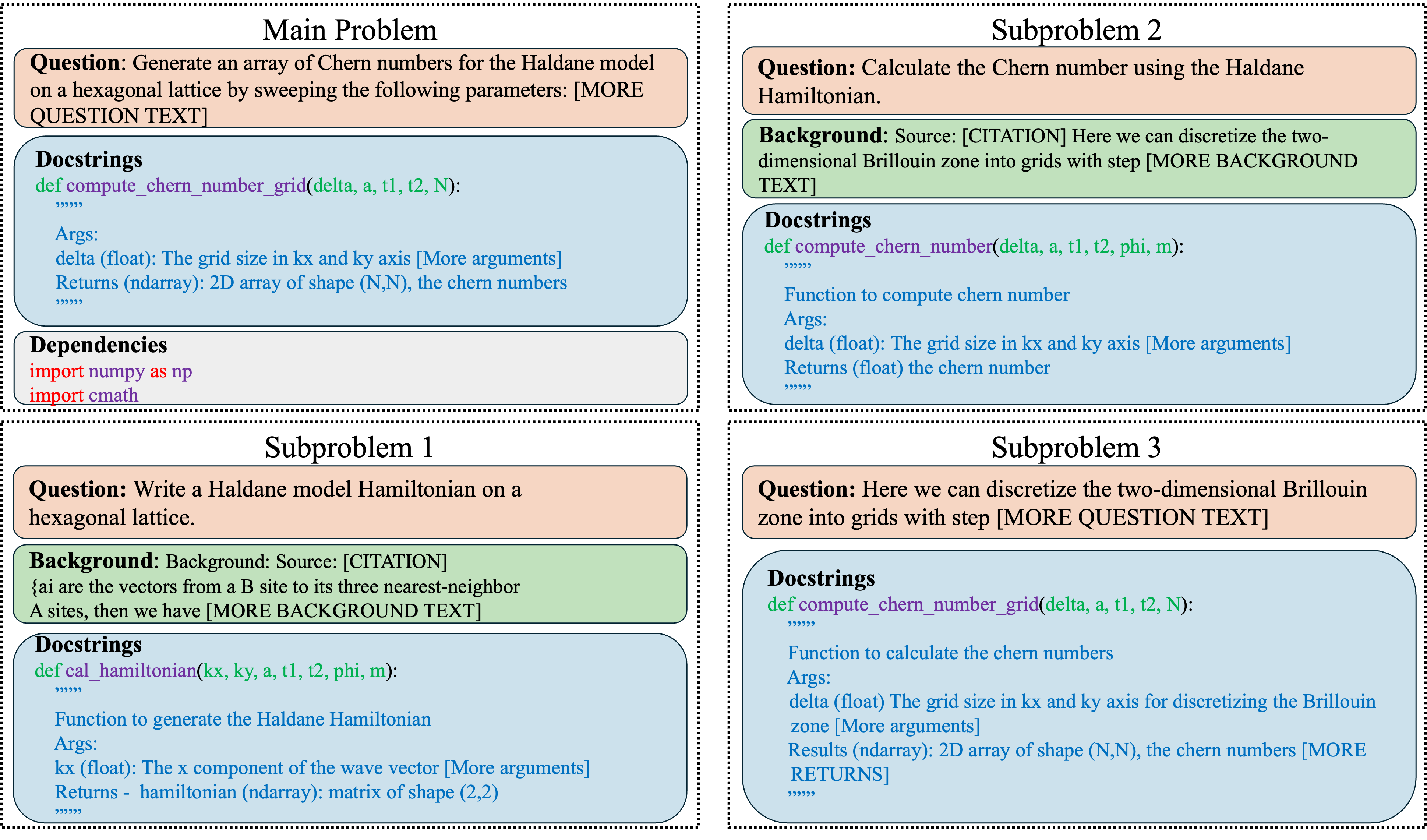}
    \caption{Structure of problems and subproblems in the SciCode dataset. Each main problem is composed of multiple subproblems, all of which must be solved correctly for the main problem to be considered successfully solved.}
    \label{fig:scicode_ds_struct}
\end{figure*}

We evaluate MOSAIC on the challenging SciCode dataset~\cite{scicode}, which covers five domains (Physics, Chemistry, Biology, Mathematics, and Materials Science) and comprises 65 main problems split into 283 subproblems. Each subproblem provides a prompt, background context, a function signature for code generation (as shown in Figure~\ref{fig:scicode_ds_struct}, Appendix Figure~\ref{appendix-fig:input-prompt-full}) without access to the gold standard ground truth code. The dataset provides a validation set of 15 main problems and 50 subproblems with ground truth code that is disjoint with the test set. We compare MOSAIC against four baseline methods: Direct, Chain of Thought (CoT), Self Planning, and Analogical. Because current state-of-the-art LLM coding frameworks rely on sample test cases to generate code and do not support integrating multiple subproblems into a unified workflow, they cannot be evaluated directly on SciCode. The test suite provided by SciCode~\cite{scicode} evaluates the correctness of a problem if all the subproblems are executable and match the ground truth target value.  To demonstrate MOSAIC’s versatility, we also evaluate it alongside leading multi-agent coding frameworks on the general-purpose MBPP dataset~\cite{MBPP} (1,000 problems) and the HumanEval dataset~\cite{humaneval} (164 problems), as well as on the APPS benchmark~\cite{APPS}, which comprises over 5,000 problems spanning introductory, interview, and competition-level challenges. On the general purpose coding datasets, we compare MOSAIC against MapCoder~\cite{mapcoder} and CodeSIM~\cite{codesim} where it achieves comparable performance.

\subsection{Implementation Details}

We build on the open-source PyTorch implementation of SciCode~\cite{scicode} for our experiments. Within MOSAIC, we employ LangGraph~\cite{langgraph} to orchestrate agent communication and ensure reproducibility. For each problem domain, we have a dedicated memory for the agentic framework to ensure encountering only the domain knowledge and prevent cross-domain interference. Each agent (Self-Reflection, Rationale, Coding, and Debugging) is guided by tailored prompts that constrain it to its specific role and yield the outputs needed to arrive at the final solution (Included in the Appendix~\ref{sec:appendix}). In MOSAIC’s teacher module, designed for knowledge distillation via few-shot prompting, we sample twenty problems at random from the APPS training set (seed 1993)~\cite{APPS}, use the ten MBPP problems provided for few-shot examples~\cite{MBPP}, and include five problems from the HumanEval dataset~\cite{humaneval} taken out of the entire dataset.

We evaluate our MOSAIC framework, other baselines and benchmarks using Open AI GPT-4o, Claude Sonnet 4 and Gemini 2.5 Flash. In our ablation studies~\ref{sec:ablation} we explore other open source LLM backbones. 

\subsection{Evaluation Metrics}

In this paper, we adopt the SciCode evaluation protocol~\cite{scicode} for the scientific dataset, counting solved sub-problems and main problems. We also report metrics that quantify how closely our outputs align with the reference solutions in SciCode (Figure~\ref{abl-fig:precision_diff}) in our Ablation section~\ref{sec:ablation}, even for cases that don’t fully succeed, since precision and accuracy are critical in scientific problem solving. For MBPP~\cite{MBPP}, HumanEval~\cite{humaneval}, and APPS~\cite{APPS}, we report performance as the percentage of test cases passed. 

\section{Results and Discussion}
\label{sec:results}

\begin{table*}[ht!]
\centering
\resizebox{\textwidth}{!}{%
\begin{tabular}{l|l|cc|cc|cc|cc|cc|cc}
\toprule
LLM Backbone & Methods 
  & \multicolumn{2}{c|}{Total} 
  & \multicolumn{2}{c|}{Physics} 
  & \multicolumn{2}{c|}{Chemistry} 
  & \multicolumn{2}{c|}{Biology} 
  & \multicolumn{2}{c|}{Material Science} 
  & \multicolumn{2}{c}{Mathematics} \\
\midrule
\multirow{6}{*}{GPT-4o}   
  & SciCode Baseline &  7/65 &  94/283  &   3/30 &  48/145 &   1/7 &   13/42 &   0/7 &   5/25 &   2/11&   24/50 &   1/10 &   4/24 \\
\cmidrule(l){2-14}
  & Analogical       &   1/65 &  32/283  &   1/30 &  18/145 &   0/7 &   2/42 &   0/7 &   3/25 &   0/11&   6/50 &   0/10 &   3/24 \\
\cmidrule(l){2-14}
  & CoT       &   2/65 &  38/283  &   1/30 &  21/145 &   0/7 &   2/42 &   0/7 &   3/25 &   1/11&   8/50 &   0/10 &   4/24 \\
\cmidrule(l){2-14}
  & LATS             &   4/65 &  49/283  &   2/30 &  34/145 &   0/7 &   2/42 &   0/7 &   3/25 &   2/11&   8/50 &   0/10 &   2/24 \\
\cmidrule(l){2-14}
  & MOSAIC (\textit{ours})
                    &  \textcolor{Cerulean}{\textbf{12/65}} & \textcolor{Cerulean}{\textbf{113/283}} &  \textcolor{Cerulean}{\textbf{4/30}} &  \textcolor{Cerulean}{\textbf{56/145}} &   \textcolor{Cerulean}{\textbf{2/7}} &  \textcolor{Cerulean}{\textbf{14/42}} &   \textcolor{Cerulean}{\textbf{0/7}} &   \textcolor{Cerulean}{\textbf{7/25}} &   \textcolor{Cerulean}{\textbf{3/11}} &  \textcolor{Cerulean}{\textbf{26/50}} &   \textcolor{Cerulean}{\textbf{3/10}} &   \textcolor{Cerulean}{\textbf{10/24}} \\
\midrule\midrule
\multirow{2}{*}{Claude Sonnet 4} 
  & SciCode Baseline &   9/65 &  109/283  &   4/30 &  71/145 &   1/7 &   13/42 &   1/7 &   8/25 &   2/11&   8/50 &   1/10 &   8/24 \\
\cmidrule(l){2-14}
  & MOSAIC (\textit{ours}) 
                    &   \textbf{\textcolor{Cerulean}{13/65}} &  \textbf{\textcolor{Cerulean}{118/283}}  &   \textbf{\textcolor{Cerulean}{4/30}} &  \textbf{\textcolor{Cerulean}{77/145}} &   \textbf{\textcolor{Cerulean}{2/7}} &   \textbf{\textcolor{Cerulean}{17/42}} &   \textbf{\textcolor{Cerulean}{1/7}} &   \textbf{\textcolor{Cerulean}{9/25}} &   \textbf{\textcolor{Cerulean}{3/11} }&  \textbf{ \textcolor{Cerulean}{8/50}} &   \textbf{\textcolor{Cerulean}{3/10}} &   \textbf{\textcolor{Cerulean}{8/24}} \\ \midrule \midrule
\multirow{2}{*}{Gemini 2.5 flash} 
  & SciCode Baseline &   7/65 &  112/283  &   5/30 &  67/145 &   1/7 &   6/42 &   1/7 &   5/25 &   2/11&   6/50 &   1/10 &   1/24 \\
\cmidrule(l){2-14}
  & MOSAIC (\textit{ours}) 
                    &   \textbf{\textcolor{Cerulean}{11/65}} &  \textbf{\textcolor{Cerulean}{117/283}}  &   \textbf{\textcolor{Cerulean}{5/30}} &  \textbf{\textcolor{Cerulean}{88/145}} &   \textbf{\textcolor{Cerulean}{2/7}} &   \textbf{\textcolor{Cerulean}{14/42}} &   \textbf{\textcolor{Cerulean}{1/7}} &   \textbf{\textcolor{Cerulean}{11/25}} &   \textbf{\textcolor{Cerulean}{2/11}} &   \textbf{\textcolor{Cerulean}{9/50}} &   \textbf{\textcolor{Cerulean}{1/10}} &   \textbf{\textcolor{Cerulean}{12/24}} \\
                     
\bottomrule
\end{tabular}%
}
\caption{Performance comparison between baselines and MOSAIC on scientific datasets with different LLM backbones. \textcolor{Cerulean}{\textbf{Best}} results are highlighted. The SciCode benchmark consists of 65 main problems comprising a total of 283 subproblems spanning physics, chemistry, biology, materials science, and mathematics. A problem is considered solved only when all of its subproblems pass the corresponding test suites.}
\label{tab:main_comp}
\end{table*}

\begin{figure*}[!h]
    \centering
    \includegraphics[width=1\textwidth]{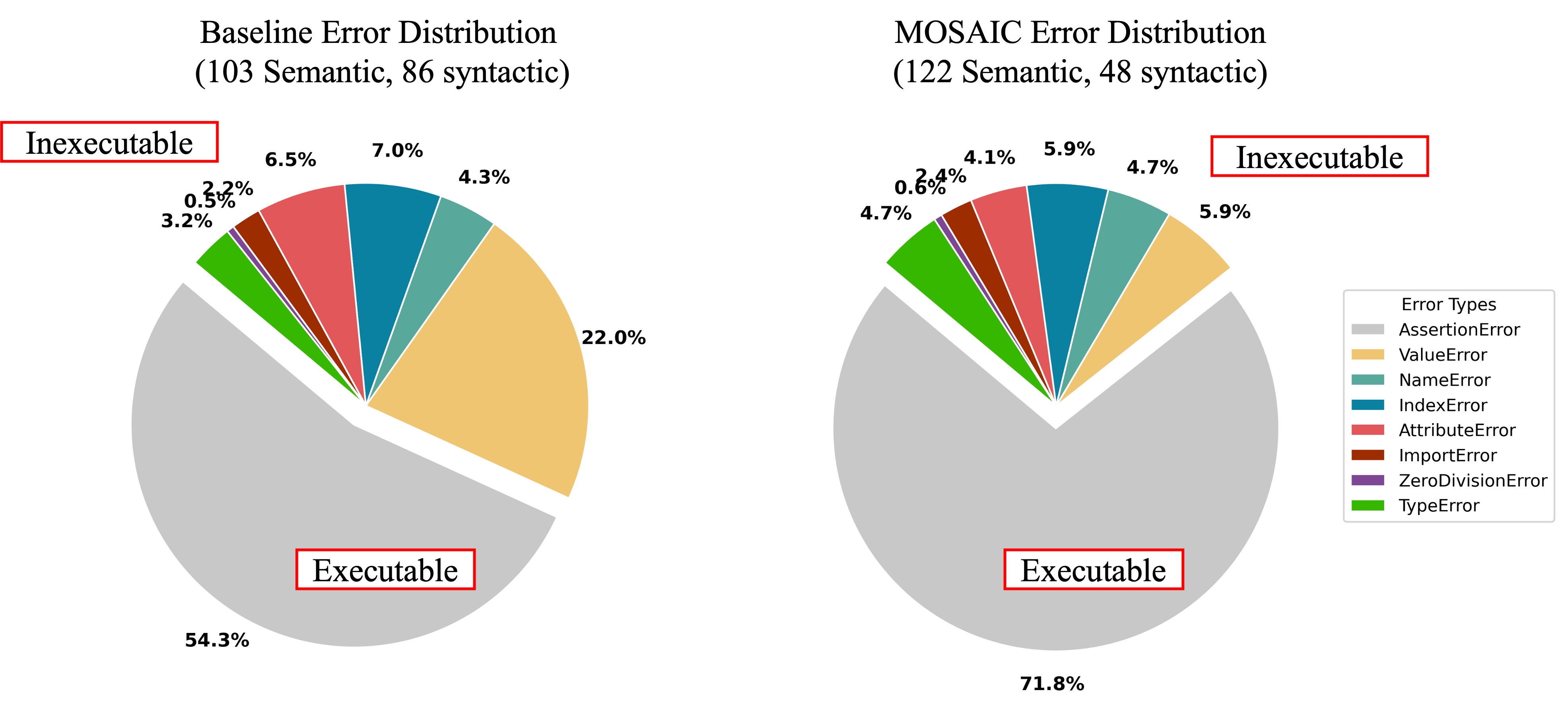}
    \caption{Error statistics on SciCode benchmarks. The figure distinguishes \textbf{syntactic errors} (failed execution) from \textbf{semantic errors} (output–target mismatches). Semantic errors are shown in gray, while other colors represent different categories of syntactic errors. MOSAIC substantially reduces both the overall error rate and the relative proportion of syntactic errors compared to the baseline.}
    \label{fig:error_types}
\end{figure*}

\begin{table*}[h!]
\centering
\resizebox{0.7\textwidth}{!}{%
\begin{tabular}{c|c|c|c}
\hline
\multirow{2}{*}{Method} &
\multirow{2}{*}{\begin{tabular}[c]{@{}c@{}}HumanEval\\ (30 val / 134 test)\end{tabular}} &
\multirow{2}{*}{\begin{tabular}[c]{@{}c@{}}MBPP\\ (10 val / 500 test)\end{tabular}} &
\multicolumn{1}{c}{APPS} \\ 
 &  &  &  \\ \hline
Direct         & 89.63              & 48.80              & 12.70 \\ \hline
CoT            & 87.20              & 54.92              & 11.30 \\ \hline
Self-Planning  & 89.63              & 49.64              & 14.70 \\ \hline
Analogical     & 90.85              & 49.81              & 12.00 \\ \hline
MapCoder       & 90.26              & 77.92              & 22.37 \\ \hline
CodeSIM        & \textcolor{Cerulean}{\textbf{93.60}} & \textcolor{Bittersweet}{\textbf{80.49}} & \textcolor{Bittersweet}{\textbf{22.56}} \\ \hline
MOSAIC (\textit{ours}) & \textcolor{Bittersweet}{\textbf{92.53}} & \textcolor{Cerulean}{\textbf{84.90}} & \textcolor{Cerulean}{\textbf{24.71}} \\ \hline
\end{tabular}%
}
\caption{Performance comparison of different code generation methods on HumanEval, MBPP, and APPS benchmarks. Reported values represent accuracy scores (\%). The \textcolor{Cerulean}{\textbf{Best}} results for each benchmark are highlighted in blue, while the \textcolor{Bittersweet}{\textbf{second best}} results are highlighted in orange. MOSAIC achieves competitive performance with CodeSIM, ranking first on MBPP and APPS and second on HumanEval}
\label{tab:gen_pur_comp}
\end{table*}

Table~\ref{tab:main_comp} reports performance across five scientific domains and three LLM backbones. MOSAIC consistently outperforms all baselines under every backbone. With GPT-4o, it solves 12/65 main problems and 113/283 subproblems, with strongest gains in Physics (56/145) and Material Science (26/50). Claude Sonnet~4 yields the best overall result (13/65, 118/283), with clear improvements in Physics (77/145) and Chemistry (17/42). Gemini 2.5 Flash achieves 11/65 and 117/283, with the largest gains in Physics (88/145) and Mathematics (12/24). Across all settings, MOSAIC's gains stem from its multi-agent orchestration: the Rationale Agent proposes a structured plan guided by the Teacher Module through domain-specific few-shot examples, while the CCW reduces hallucination by exposing only prior function signatures and one-line summaries to each subsequent agent.
 
Performance in Biology is consistently the weakest across all backbones. We observed incorrect ordering of steps and oversimplified algorithmic logic errors that are very different from numerical precision failures in physics or mathematics. We identified three factors that contribute to this lower performance. (1) LLMs struggle
to transfer concepts across structurally related prompts~\cite{xu2024llmsabstractionreasoningcorpus} (2)
biological knowledge in closed-source models is difficult to verify and (3) Biology is the least represented domain in the SciCode validation set.
Which leaves fewer ground-truth samples for creating guidance templates to use for few-shot prompting. For problems
with more than 10 subproblems, even the CCW is insufficient to maintain
full context the model does not consistently reuse prior function headers or carry forward intermediate results.
 
\paragraph{LLM backbone insights}
Different backbones excel in different roles, Claude Sonnet~4 performs
better on code generation tasks, while Gemini provides stronger reasoning and GPT-4o does moderately well on both. This points to the potential of heterogeneous agent configurations, which
we leave for future work. We also found that capitalizing critical
instructions in prompts (e.g., \texttt{DO NOT}) reliably kept agents within intended bounds, consistent with tokenization differences between \texttt{text} and \texttt{TEXT}.
 
\begin{figure}[h!]
    \centering
    \includegraphics[width=1\linewidth]{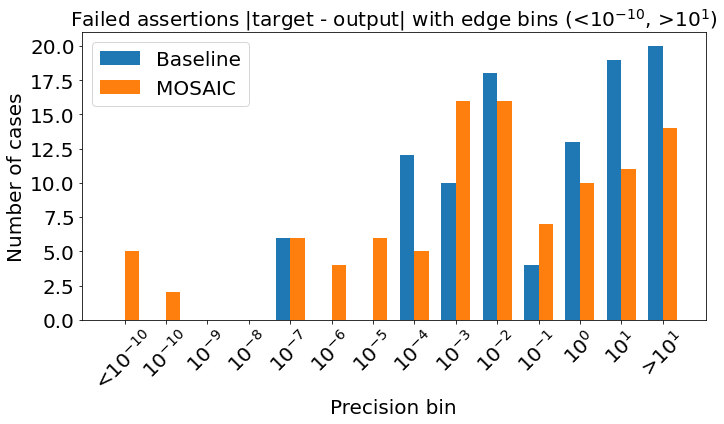}
    \caption{\textbf{Precision differences between target and generated outputs.} Compared to the baseline, MOSAIC produces a larger proportion of executable code, which introduces slightly more detectable errors. However, \textbf{MOSAIC outputs exhibit substantially smaller deviations} from the target values, indicating improved numerical precision.}
    \label{abl-fig:precision_diff}
\end{figure}

\paragraph{Error analysis.}
Figure~\ref{fig:error_types} shows the distribution of syntactic and semantic errors across benchmarks. The baseline produces nearly half its programs with syntactic errors, rendering them inexecutable. MOSAIC shifts this distribution, the overall error count drops, and the remaining errors are predominantly semantic (AssertionError), meaning the code runs but produces an incorrect value. This shift is deliberate. MOSAIC's Debugger Agent is constrained to syntactic grounding only, so semantic errors surface as a natural consequence of ensuring executability first. Once execution is guaranteed, algorithmic refinement becomes easier. Figure~\ref{abl-fig:precision_diff} further shows that MOSAIC outputs have substantially smaller deviations from target values, reflecting improved numerical precision across domains.
 
\paragraph{General coding benchmarks.}
Despite being designed for I/O-free scientific settings, MOSAIC also
performs competitively on standard benchmarks (Table~\ref{tab:gen_pur_comp}), ranking first on MBPP (84.90\%) and APPS (24.71\%) and second on HumanEval (92.53\%), behind CodeSIM. This confirms that architectural grounding does not compromise general coding capability.
\section{Ablation Studies}
\label{sec:ablation}

\begin{table*}[!h]
\resizebox{1\textwidth}{!}{%
\begin{tabular}{l|c|c|c|c|c|c}
\hline
Method           & Total        & Phys prob    & Chem     & Biology & Mat Sci & Math     \\ \hline
Baseline & 7/65, 94/283 & 3/30, 48/145 & 1/7, 13/42 & 0/7, 5/25 & 2/11, 24/50 & 1/10, 4/24 \\ \midrule
\begin{tabular}[c]{@{}l@{}}Baseline + Rationale +\\ Coding and Debug Agent\end{tabular} &
  9/65, 97/283 &
  3/30, 47/145 &
  2/7, 15/42 &
  0/7, 6/25 &
  3/11, 25/50 &
  1/10, 4/24 \\ \midrule
\begin{tabular}[c]{@{}l@{}}Baseline + Few-shot prompting + Rationale +\\ CCW(all prev. code) + Coding and Debug Agent\end{tabular} &
    4/65, 57/283 &
  1/30, 32/145 &
  1/7, 6/42 &
  0/7, 6/25 &
  1/11, 9/50 &
  1/10, 4/24 \\ \midrule
  \begin{tabular}[c]{@{}l@{}}Baseline + Self Reflection (stepwise) \\ + Few-shot prompting + Rationale \\+ CCW(prev. headers) + Coding and Debug Agent\end{tabular} &
  {6/ 65, 81/283} &
  {2/30, 48/145} &
  {1/7,  8/42} &
  {0/7, 6/25} &
  {2/11, 14/50} &
  {0/10, 5/24} \\ \midrule
  \begin{tabular}[c]{@{}l@{}}\textbf{MOSAIC(\textit{ours})} = \\ Baseline + Self Reflection (whole) \\ + Few-shot prompting + Rationale \\+ CCW(prev. headers) + Coding and Debug Agent\end{tabular} &
  \textbf{12/ 65, 113/283} &
  \textbf{4/30, 56/145} &
  \textbf{2/7,  14/42} &
  \textbf{0/7, 7/25} &
  \textbf{3/11, 26/50} &
  \textbf{3/10, 10/24} \\ \bottomrule
\end{tabular}
}
\caption{Performance comparison of MOSAIC with incremental addition of specialized agents (Rationale, Coding, Debugger) and mechanisms (Consolidated Context Window (CCW), Self-Reflection, and Few-shot prompting). The benchmark baseline~\cite{scicode} is included for reference. The results highlight how each component contributes to overall performance, with the full MOSAIC framework yielding the most significant improvements.}
\label{tab:abl_parts_comp}
\end{table*}


\begin{table*}[t!]
\centering
\begin{tabular}{@{}l|l|l|l|cc}
\toprule
Method & Type          & LLM backbone     & Parameters & \multicolumn{2}{c}{Problems solved}      \\ \cmidrule(l){5-6} 
       &               &                  &            & \multicolumn{1}{c|}{Main} & Sub Problems \\ \midrule
\multirow{8}{*}{MOSAIC(\textit{ours})}       & Open Source   & Mistral          & 7B         & \multicolumn{1}{l|}{0/65}     & 24/283              \\
       & Open Source   & Gemma 3          & 27B        & \multicolumn{1}{l|}{1/65}     &  39/283              \\
       & Open Source   & Llama 4          & 16$\times$17B       & \multicolumn{1}{l|}{2/65}     & 41/283             \\
       & Open Source   & DeepSeek R1      & 32B       & \multicolumn{1}{l|}{4/65}     & 84/283              \\ \cmidrule(l){2-6} 
       & Closed Source & Gemini 2.5 Flash & NA         & \multicolumn{1}{l|}{11/65}     & 117/283              \\
       & Closed Source & Claude Sonnet 4 & NA         & \multicolumn{1}{l|}{13/65}     & 118/283             \\
       & Closed Source & GPT-4o           & NA         & \multicolumn{1}{l|}{12/65}     & 113/283             \\ \bottomrule
\end{tabular}
\caption{Performance comparison of Open and Closed Source models on SciCode benchmark dataset. \textbf{Open-source models solve at most 4/65 main problems and 84/283 subproblems} (DeepSeek R1), while smaller ones like Mistral fail to solve any main problems. \textbf{Closed-source models perform substantially better}, with Claude Sonnet 4 achieving 13/65 main and 118/283 subproblems solved. On average, \textbf{closed-source} backbones solve nearly \textbf{3× more main problems} and \textbf{2× more subproblems}, highlighting the current performance gap.}
\label{tab:abl_cl_op_model}
\end{table*}


\subsection{Component Analysis}
 
Table~\ref{tab:abl_parts_comp} reports results as agents and mechanisms are added incrementally to the SciCode baseline (GPT-4o). Adding a Rationale Agent with iterative Coding and Debugger Agents improves performance from 7/65 to 9/65 main problems (+4\% solve rate), primarily by reducing syntactic errors. Adding few-shot prompting alongside an unrestricted CCW (retaining all prior code) surprisingly \emph{drops} performance to 4/65, as the expanded context caused the model to replicate irrelevant code fragments and produce logically inconsistent outputs. Restricting the CCW to function signatures and one-line summaries recovers most of this loss
but still lacks a mechanism to generalize reasoning across problems.
 
Introducing a Self-Reflection Agent that processes rationales
\emph{step-by-step} gives 6/65 still below baseline because
fragmenting the rationale loses overall problem context. Applying
self-reflection over the \emph{entire} rationale preserves this context
and yields best performance with 12/65 main problems solved and 113/283 subproblems solved, an 8.5\% gain over baseline. Two key insights emerge. First, components added in isolation can \emph{hurt} performance and careful orchestration mattered more than component count. Second, each configuration solves a somewhat different subset of problems; because overlap is minimal, combining all agents broadens coverage and gives the largest overall gain.
 
\subsection{Open- vs.\ Closed-Source Models}
 
Table~\ref{tab:abl_cl_op_model} compares closed-source and open-source
backbones. Among open-source models, DeepSeek R1 (32B) performs best
(4/65, 84/283), but does not surpass the SciCode baseline with GPT-4o.
Mistral 7B solves no main problems. Closed-source models substantially
outperform open-source counterparts on average solving $3\times$ more
main problems and $2\times$ more subproblems which we attribute to larger and more diverse training data, proprietary fine-tuning, and stronger alignment. Domain-specific fine-tuning of open-source models on curated scientific data is a promising direction to close this gap.
 
\subsection{Numerical Precision}
 
Figure~\ref{abl-fig:precision_diff} reports the distribution of output deviations from target values (binned from $<10^{-10}$ to $>10^1$). MOSAIC consistently reduces large deviations, producing outputs that are both more precise and better aligned with the underlying problem structure. In scientific and engineering contexts, small numerical errors accumulate and can substantially affect downstream tasks. The improvement is a direct consequence of the student-teacher structure, which ensures domain specific guidance is provided through structured templates to help the Rationale Agent refine it's plan before code generation. 
\section{Conclusion}
\label{sec:conclusion}
Scientific code generation is inherently more challenging than general-purpose coding, as there are no I/O test cases for validation, and generating such test cases itself requires solving the underlying problem. Instead of relying entirely on execution-based feedback for verifying algorithmic correctness, our solution MOSAIC adopts a student–teacher distillation framework to guide code generation through domain-specific reasoning. Extensive experiments and ablation studies on the SciCode benchmark demonstrate the effectiveness and robustness of our approach, achieving up to 24\% higher accuracy even with lightweight models. Overall, our results highlight the practical utility of MOSAIC and provide a strong foundation for future research in scientific code generation. 

\bibliography{custom}

\clearpage
\appendix
\section{Appendix}
\label{sec:appendix}

\begin{figure*}[t]
    \centering
    \includegraphics[width=\linewidth]{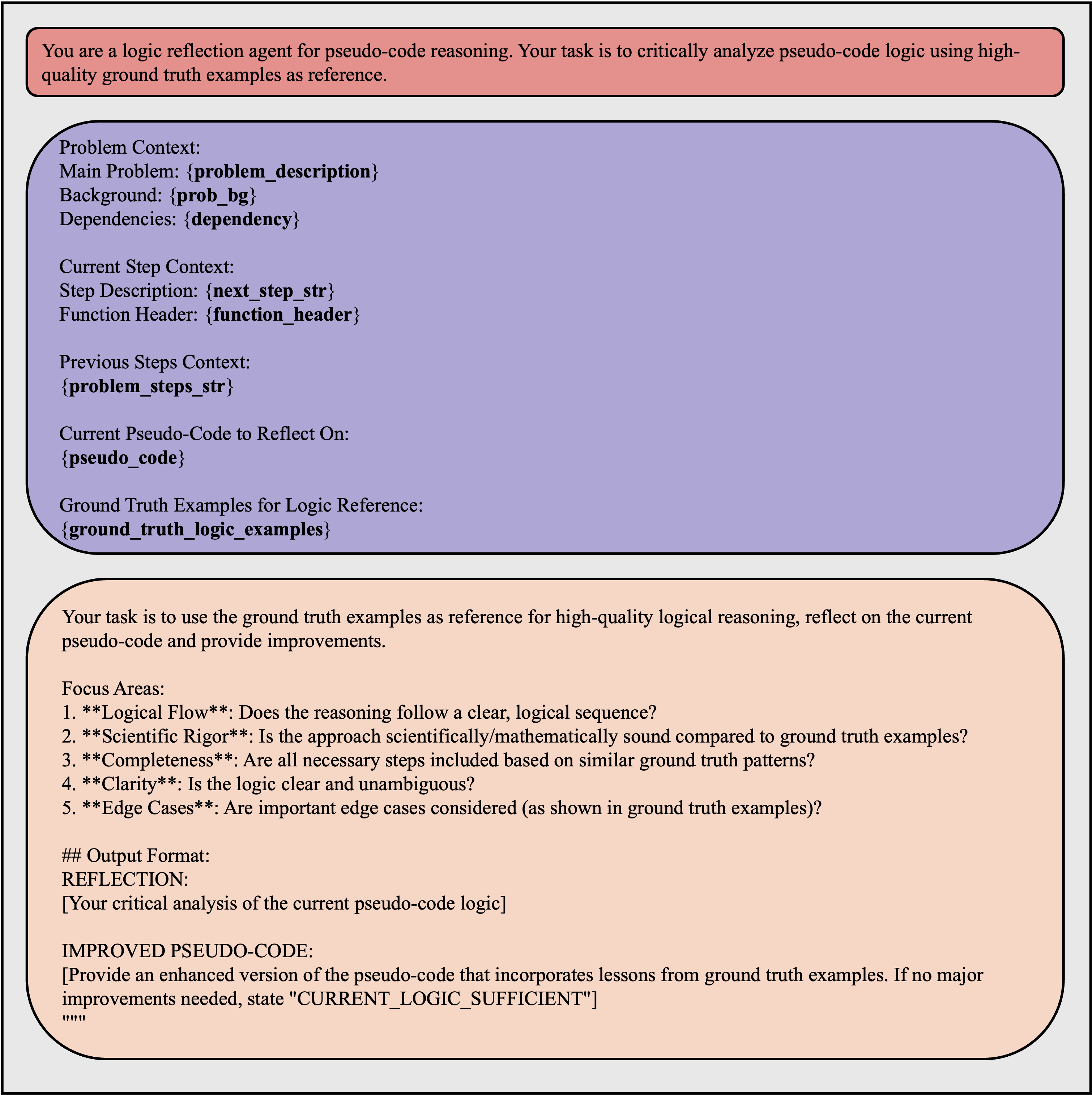}
    \caption{Prompt template for the \textbf{Self-Reflection Agent}, which
    analyzes ground-truth code from the validation set to identify
    domain-specific patterns and prepares gold-standard pseudocode as
    few-shot examples for the Rationale Agent.}
    \label{appendix_fig:reflect-prompt}
\end{figure*}

\begin{figure*}[t]
    \centering
    \includegraphics[width=\linewidth]{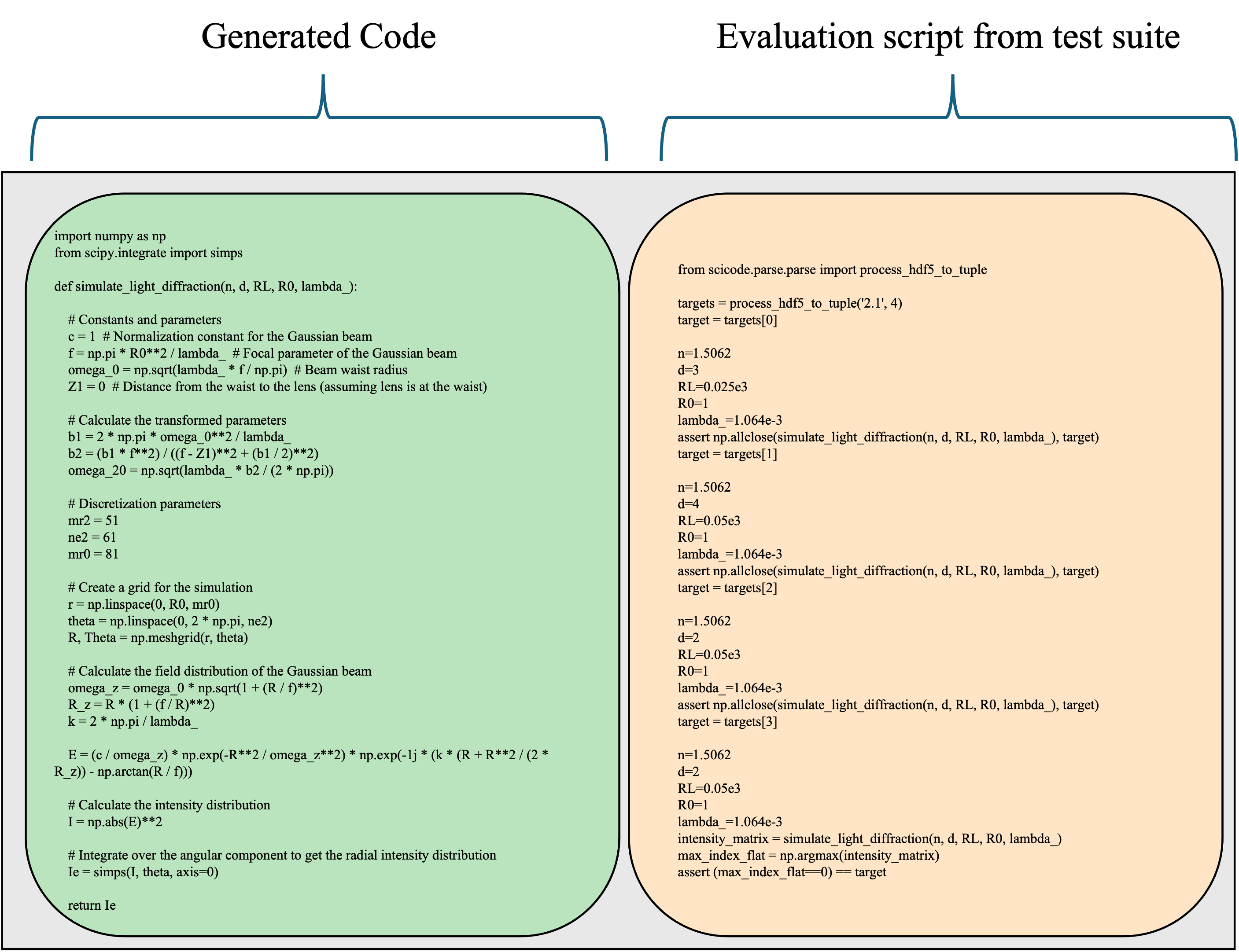}
    \caption{A sample of code generated by MOSAIC, alongside the test suite
    evaluation code used to verify correctness.}
    \label{appendix_fig:code-generated}
\end{figure*}

In the Appendix, we illustrate the overall format of the SciCode benchmark input prompt (Figure~\ref{appendix-fig:input-prompt-full}), present the code generated by the SciCode baseline and MOSAIC side by side (Figure~\ref{appendix-code:comparison}), and provide the benchmark's test-suite snippet and error traceback
(Figure~\ref{appendix-code:test-suite}). We also provide the complete set of prompt templates used in MOSAIC
(Figures~\ref{appendix_fig:reflect-prompt}, \ref{appendix_fig:code-generated}). We also discuss some key limitations of our work.

\subsection{Limitations}

MOSAIC has two main limitations. First, it relies on the domain labels provided by SciCode rather than inferring them automatically. When we tested LLM-based domain bucketing using domain-specific keywords, the model frequently assigned problems to the wrong domain, leading to a performance drop of around 10 to 12\%. Second, since MOSAIC is training-free and LLM-agnostic, its performance is bounded by the capabilities of the underlying backbone. As shown in Main paper Table~\ref{tab:abl_cl_op_model}, open-source models lag significantly behind closed-source ones on SciCode. Domain-specific fine-tuning of open-source models and heterogeneous agent configurations, where different models specialize in different roles, are promising directions for future work.
\subsection{Code Outputs}

Each problem in SciCode may contain several subproblems, and a problem is considered complete only when all subproblems are solved without errors. Every subproblem includes a step description, a set of inputs, an expected output type, and a function header that the agent must follow precisely. A central challenge of SciCode is the absence of sample inputs and outputs for checking intermediate correctness; the agent has access only to domain-level validation examples that differ from the problem in the input prompt (Figure~\ref{appendix-fig:input-prompt-full}).

The SciCode test suite calls \texttt{ket(2, [1,1])}
(Figure~\ref{appendix-code:test-suite}), intended to construct the tensor product state $\lvert 1 \rangle \otimes \lvert 1 \rangle$, whose expected output is the four-dimensional column vector $[0, 0, 0, 1]^{T}$. The baseline does not distinguish between the case where \texttt{dim} is an integer and \texttt{args} is a list, treating the list as a NumPy fancy indexing operation and producing a vector of shape $(2,1)$ instead of $(4,1)$. NumPy cannot broadcast these shapes for element-wise comparison, causing a runtime error. This is an algorithmic error rather than a syntactic one. MOSAIC instead constructs each basis vector separately and computes their Kronecker product, producing the correct shape $(4,1)$ that satisfies the test without error.

\begin{figure*}[t]
\centering
\begin{minipage}{0.48\linewidth}
\begin{lstlisting}[language=Python, basicstyle=\footnotesize\ttfamily, numbers=none]
from parse.parse import process_hdf5_to_tuple

targets = process_hdf5_to_tuple('11.1', 3)
target = targets[0]
assert np.allclose(ket(2, 0), target)
target = targets[1]
assert np.allclose(ket(2, [1,1]), target)
target = targets[2]
assert np.allclose(ket([2,3], [0,1]), target)
\end{lstlisting}
\end{minipage}
\hfill
\begin{minipage}{0.48\linewidth}
\begin{lstlisting}[language=Python, basicstyle=\footnotesize\ttfamily, numbers=none]{text}
Traceback (most recent call last):
  File "main.py", line 42, in <module>
    assert np.allclose(ket(2, [1,1]), target)
  ...
ValueError: operands could not be
broadcast together with shapes
(2,1) (4,1)
\end{lstlisting}
\end{minipage}
\vspace{0.5em}
\begin{minipage}{0.48\linewidth}
  \caption*{\textbf{(a)} Test suite snippet used to evaluate correctness.}
\end{minipage}
\hfill
\begin{minipage}{0.48\linewidth}
  \caption*{\textbf{(b)} Python traceback when running the SciCode baseline.}
\end{minipage}
\caption{SciCode evaluation: test suite call and the resulting baseline error.}
\label{appendix-code:test-suite}
\end{figure*}

\begin{figure*}[t]
\centering
\begin{minipage}{0.48\linewidth}
\begin{lstlisting}[language=Python, basicstyle=\footnotesize\ttfamily, numbers=none]
def ket(dim, args):
    '''Input:
    dim: int or list, dimension of the ket
    args: int or list, the i-th basis vector
    Output:
    out: dim dimensional array of float,
         the matrix representation of the ket
    '''
    if isinstance(dim, int):
        out = np.zeros(dim)
        out[args] = 1.0
    else:
        basis_vectors = []
        for d, j in zip(dim, args):
            vec = np.zeros(d)
            vec[j] = 1.0
            basis_vectors.append(vec)
        out = basis_vectors[0]
        for vec in basis_vectors[1:]:
            out = np.kron(out, vec)
    return out.reshape(-1, 1)
\end{lstlisting}
\end{minipage}
\hfill
\begin{minipage}{0.48\linewidth}
\begin{lstlisting}[language=Python, basicstyle=\footnotesize\ttfamily, numbers=none]
def ket(dim, args):
    '''Input:
    dim: int or list, dimension of the ket
    args: int or list, the i-th basis vector
    Output:
    out: dim dimensional array of float,
         the matrix representation of the ket
    '''
    if isinstance(dim, int) and isinstance(args, int):
        out = np.zeros(dim)
        out[args] = 1.0
        return out.reshape(-1, 1)
    elif isinstance(dim, int) and isinstance(
                             args, (list, tuple)):
        basis_vectors = []
        for j in args:
            vec = np.zeros(dim)
            vec[j] = 1.0
            basis_vectors.append(vec)
        out = basis_vectors[0]
        for vec in basis_vectors[1:]:
            out = np.kron(out, vec)
        return out.reshape(-1, 1)
    else:
        basis_vectors = []
        for d, j in zip(dim, args):
            vec = np.zeros(d)
            vec[j] = 1.0
            basis_vectors.append(vec)
        out = basis_vectors[0]
        for vec in basis_vectors[1:]:
            out = np.kron(out, vec)
        return out.reshape(-1, 1)
\end{lstlisting}
\end{minipage}
\vspace{0.5em}
\begin{minipage}{0.48\linewidth}
  \caption*{\textbf{(a)} SciCode baseline~\cite{scicode}: fails to handle
  the case where \texttt{dim} is int and \texttt{args} is a list.}
\end{minipage}
\hfill
\begin{minipage}{0.48\linewidth}
  \caption*{\textbf{(b)} MOSAIC \textit{(ours)}: correctly computes the
  Kronecker product for all input combinations.}
\end{minipage}
\caption{Side-by-side comparison of the SciCode baseline and MOSAIC outputs
for the \texttt{ket} function.}
\label{appendix-code:comparison}
\end{figure*}

\begin{figure*}[t]
\centering
\begin{minipage}{0.95\linewidth}
\begin{tcolorbox}[title=Input prompt]
\textbf{Main problem:}
Consider sending a bipartite maximally entangled state where both parties
are encoded by $m$-rail encoding through $m$ uses of the generalized
amplitude damping channel $\mathcal{A}_{\gamma_1,N_1}$ to receiver~1 and
$m$ uses of another generalized amplitude damping channel
$\mathcal{A}_{\gamma_2,N_2}$ to receiver~2. Each of the two receivers
measures whether the $m$ qubits are in the one-particle sector, i.e.,
whether there are $m-1$ zeros and one excitation. If so, they keep the
state. Otherwise, they discard the state. They then perform the hashing
protocol on the post-selected state. Calculate the rate of entanglement
that can be generated per channel use in this set-up.

\medskip
\textbf{Step description:}
Given $j$ and $d$, write a function that returns a standard basis vector
$\lvert j\rangle$ in $d$-dimensional space. If $d$ is given as an int and
$j$ is given as a list $[j_1,j_2,\dots,j_n]$, return the tensor product
$\lvert j_1\rangle \lvert j_2\rangle \cdots \lvert j_n\rangle$ of
$d$-dimensional basis vectors. If $d$ is also a list $[d_1,d_2,\dots,d_n]$,
return the tensor product of $d_1, d_2, \dots, d_n$-dimensional basis
vectors.

\medskip
\textbf{Inputs:}
\begin{itemize}[noitemsep, topsep=2pt]
    \item \texttt{rails}: int, number of rails
    \item \texttt{$\gamma_1$}: float, damping parameter of the first channel
    \item \texttt{$N_1$}: float, thermal parameter of the first channel
    \item \texttt{$\gamma_2$}: float, damping parameter of the second channel
    \item \texttt{$N_2$}: float, thermal parameter of the second channel
\end{itemize}
\textbf{Output:}
\begin{itemize}[noitemsep, topsep=2pt]
    \item float, the achievable rate of our protocol
\end{itemize}

\medskip
\textbf{Function header:}
\begin{tcolorbox}[colback=black!5!white, colframe=black!40]
\texttt{def ket(dim):}\\
\texttt{\ \ \ \ '''Input:}\\
\texttt{\ \ \ \ \ \ dim: int or list, dimension of the ket}\\
\texttt{\ \ \ \ \ \ args: int or list, the i-th basis vector}\\
\texttt{\ \ \ \ Output:}\\
\texttt{\ \ \ \ \ \ out: dim dimensional array of float}\\
\texttt{\ \ \ \ '''}
\end{tcolorbox}
\end{tcolorbox}
\end{minipage}
\caption{Input prompt containing the main problem, step description, and
function header.}
\label{appendix-fig:input-prompt-full}
\end{figure*}

\subsection{Prompt Templates}

\begin{figure*}[t]
    \centering
    \includegraphics[width=\linewidth]{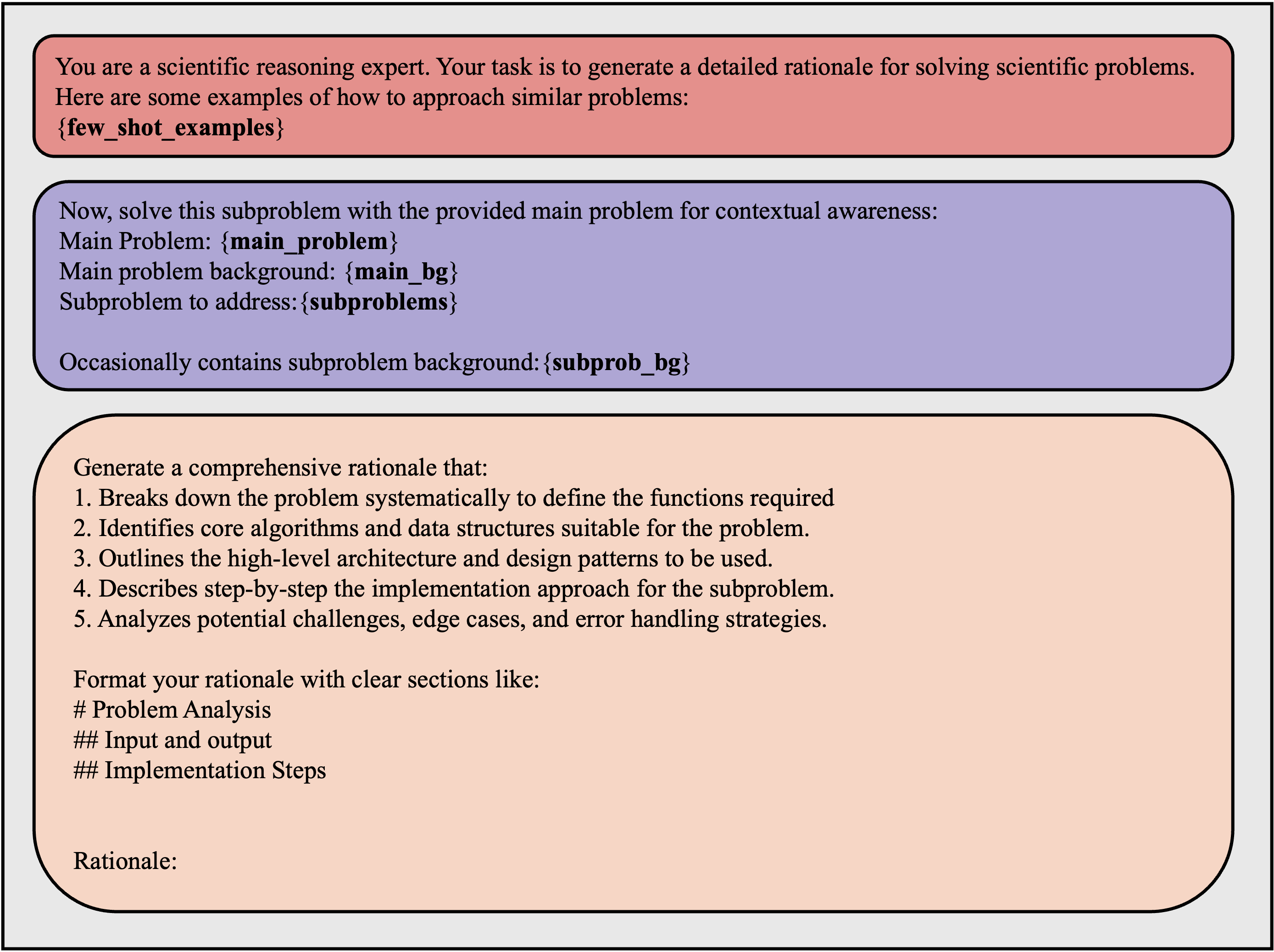}
    \caption{Prompt template for the \textbf{Rationale Agent}. Guided by
    few-shot examples from the Self-Reflection Agent, it converts the
    subproblem into a detailed rationale for the Coding Agent.}
    \label{appendix_fig:rationale-prompt}
\end{figure*}

\begin{figure*}[t]
    \centering
    \includegraphics[width=\linewidth]{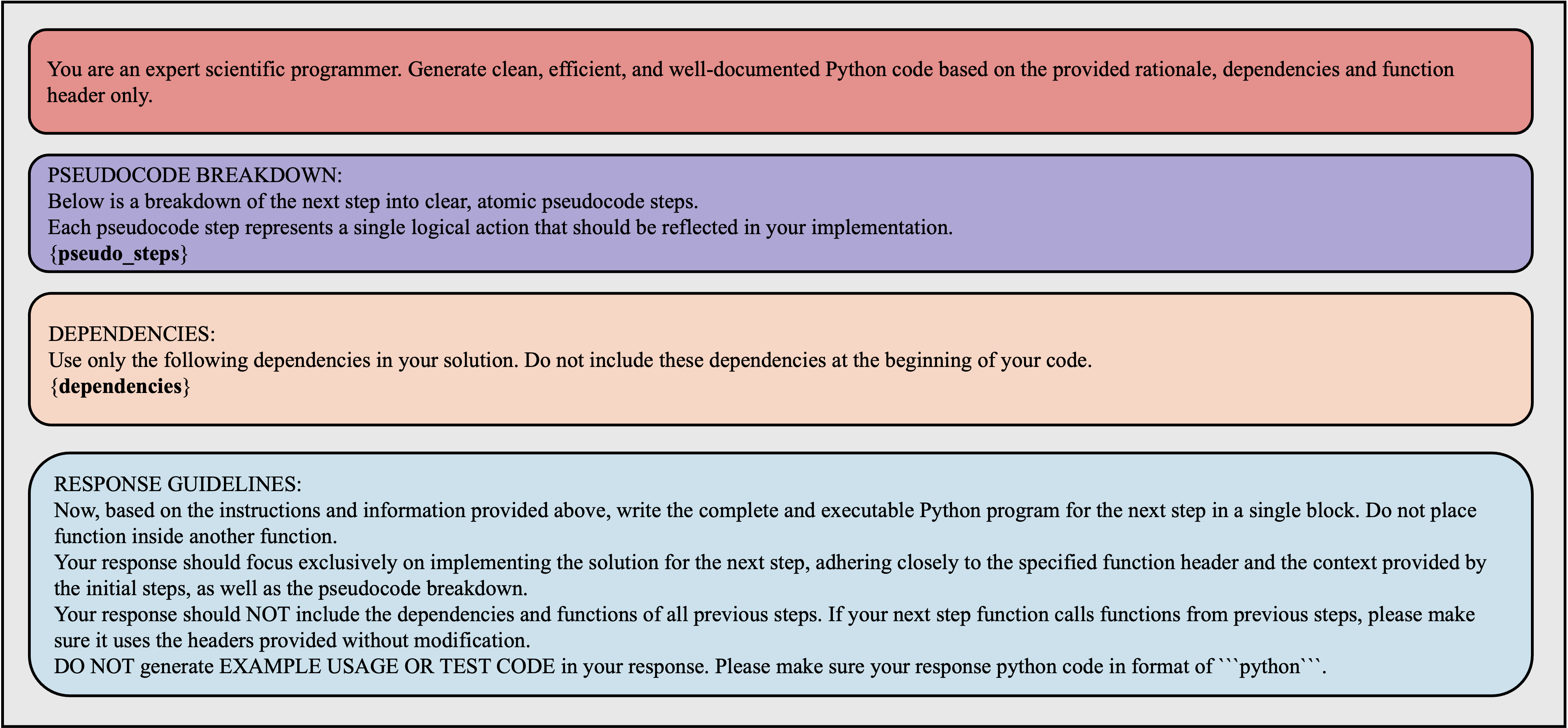}
    \caption{Prompt template for the \textbf{Coding Agent}, which converts
    each pseudocode step into executable code.}
    \label{appendix_fig:code-prompt}
\end{figure*}

\begin{figure*}[t]
    \centering
    \includegraphics[width=\linewidth]{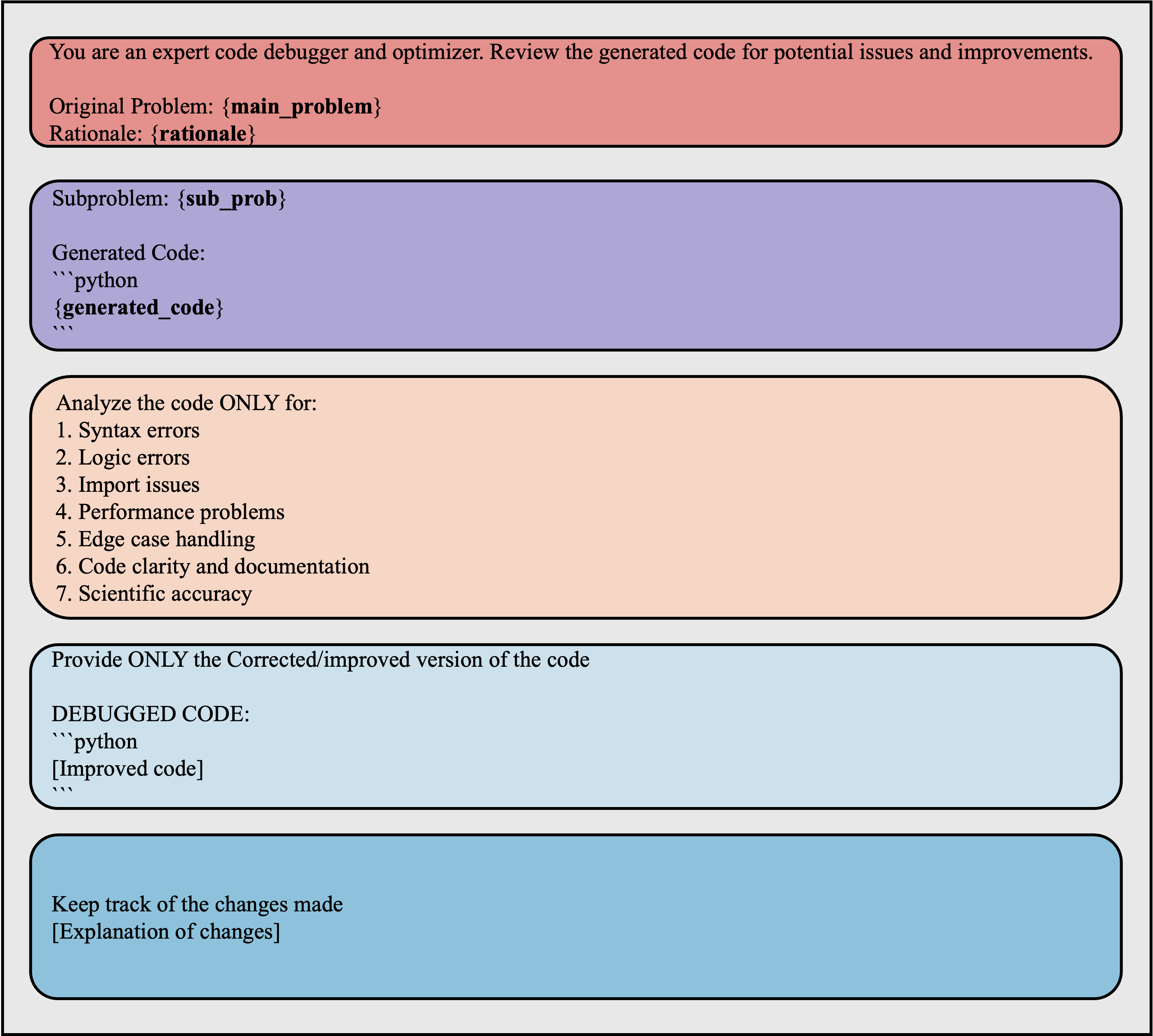}
    \caption{Prompt template for the \textbf{Debugger Agent}, which
    identifies and corrects syntactic errors to ensure executability of the
    generated code.}
    \label{appendix_fig:debug-prompt}
\end{figure*}

\end{document}